\documentstyle[prl,aps]{revtex}
\begin{document}
\draft
\tighten
\title{Confirmation of the Modified Bean Model from Simulations of 
Superconducting Vortices }
\author{R. A. Richardson, O. Pla\cite{nadd} and F. Nori\cite{dagger}}
\address{Physics Department, University of Michigan, Ann Arbor MI 48109-1120}
\date{\today}
\maketitle
\begin{abstract}
>From a very simple description of vortices and pinning centers
and without using any electrodynamical assumptions, we obtain nonlinear 
density profiles of vortices in type-II dirty superconductors that
result from raising or lowering an external magnetic field. The results
confirm a modified Bean model description of
these systems, and in particular, follow the Kim empirical form that
relates the current inside the material to the local magnetic field. We
also obtain realistic magnetization hysteresis loops and examine the 
discrete evolution of the density profiles in our
systems. This evolution is not continuous, but takes place by the
occurrence of avalanches of vortices promoted by the addition or
extraction of vortices from the edges of the system.
\end{abstract}
\narrowtext
\twocolumn

Over the past three decades, the Bean model \cite{bean} for the
magnetization of hard type-II superconductors in a varying external
magnetic field has been successfully applied to a variety of
experimental phenomena.  In most applications, the Bean form for the
flux density profiles within a sample can only be inferred from  bulk
quantities such as the magnetization, and details about the internal
field are hard to come by.   One means of providing specific
information about the behavior of flux lines within a material is
through computer simulation (see, {\em e.g.,}\cite{brandt2,pla}).
In this work, we simulate the development of Bean-type profiles in a
one dimensional system with a random array of pinning centers. We
examine the dependence of these profiles on various parameters and
demonstrate that the only physical elements necessary to achieve
Bean-like behavior are repulsive forces between vortices and attractive
forces between vortices and pinning centers.  We also investigate the
evolution of these profiles and find that the dynamics of vortices
within these systems obey power laws suggestive of self-organized
criticality  \cite{bak}.  The literature on vortices and the Bean model
is vast and we do not attempt a review.  The interested reader is
directed to the references \cite{tinkham,brandt1} and the papers cited therein
for a more complete discussion.

Our samples are one dimensional, with periodic boundary conditions
({\em i.e.,} the sample is a ring).  Pinning centers of uniform depth
are distributed randomly over most of the sample, with a small region
left pin-free from which vortices are either added or extracted.  This
pin-free region, as will be explained more fully below, can be thought
of as the region where the ``external field" is applied.

For the interaction between vortices and pinning centers we will
consider the latter to be represented by parabolic potentials. These
potentials have a finite range, which is given by the constant
$\xi_{\rm p}$. Explicitly, the total pinning potential acting on vortex
$i$ is:
\FL
\begin{equation}
V_{{\rm p} i}\!=\!\sum_j \!\left(\!-\frac{f_{\rm c}\xi_{\rm p}}
{2}+\frac{f_{\rm c}}{2\xi_{\rm p}}
(x^{\rm v}_i\!-\!x^{\rm p}_j)^2\!\right)\!\Theta(\xi_{\rm p}
\!-\!|x^{\rm v}_i\!-\!x^{\rm p}_j|),
\end{equation}
where $x_i^{\rm v}$ and $x_j^{\rm p}$ and represent the position of the
$i$th vortex and the  $j$th pinning site, respectively.  $\Theta(x)$ is
the Heaviside step function and $f_{\rm c}$ is the force needed to
detach one vortex from one pinning center in the one body problem, or,
in other words the strength of the pinning well.

The interaction between vortices is much better understood than that
between vortices and material defects (pinning sites). Several
theoretical methods (see for instance Ref. \cite{tinkham}) lead to a
vortex-vortex potential which varies as a modified Bessel function.
The evaluation of this potential for a system with many vortices is
very costly in terms of simulation time, and we therefore approximate
it by the following simple parabolic form:
\begin{equation}
V_{{\rm v} i}=\sum_{j,j\neq i} {\rm A}_{\rm v}(|x^{\rm v}_i\!-\!x^{\rm v}_j|
-\xi_{\rm v})^2\Theta(\xi_{\rm v}\!-\!|x^{\rm v}_i\!-\!x^{\rm v}_j|),
\end{equation}
where A$_{\rm v}$ is the strength of the potential and $\xi_{\rm v}$ is
the vortex-vortex interaction range. This range, like that for
vortex-pinning interaction, is taken to be finite.  We neglect the hard
core part in the potential, where the material is normal. This
corresponds to assuming that the Ginzburg-Landau parameter, $\kappa$ is
large.

The simulation consists of two procedures which correspond to slowly
ramping up and ramping down the external field.  Vortices are either
added or extracted and the system is then allowed to relax to
equilibrium.  If friction is neglected, the total force that acts on
vortex $i$ in the presence of other vortices and pinning centers is
$\vec{\cal F}_i=-\vec{\nabla}_iV_{{\rm p} i}-\vec{\nabla}_iV_{{\rm v}
i}$.  We exploit the fact that $\vec{\cal F}_i$ is linear in $x_i^{\rm
v}$ in 1D to calculate successive configurations of the flux lattice.
This is performed iteratively by updating $x_i^{\rm v}$ in such a way
that $\vec{\cal F}_i$ vanishes, where we take as input the $x_j^{\rm
v}$ calculated in the previous step. Throughout this work, the
dimensions of length and force will be referenced to $\xi_{\rm v}$ and
$f_{\rm c}$, i.e. $\xi_{\rm v}=1$ and $f_{\rm c}=1$. The free
parameters are thus A$_{\rm v}$, $N_{\rm p}$,  and $\xi_{\rm p}$.

The simulations are performed by randomly distributing $N_{\rm p}$
pinning centers over 80$\%$ of the system.  A 20$\%$ chunk is left
unpinned. The pinned region corresponds to the actual sample.  Within
the pinning region, the pins are distributed quite densely (with an
average density from 10 to 30 per unit length)  but the pinning range,
$\xi_{\rm p}$ is made small enough that the probability of pins
overlapping is quite low ($\xi_{\rm p} = 0.005$ for all data presented
here). This is done to avoid complicated pinning landscapes where some
local pinning regions are much deeper than others due to pin overlap.
In general, then, the pinning force in our models is provided by
narrow, weak and densely distributed pins of uniform depth.

In the ``ramp-up" phase of the simulation,  vortices are successively
added to the unpinned region and in the ``ramp-down" phase, they are
extracted from this region.  First, let us consider the ramp-up phase.
The first few vortices added to the system remain in the unpinned
region (UR) until, after a time, enough vortices have been added that
they cannot all remain in the unpinned region without overlapping.  At
this point,  vortices begin to be forced into the pinned region (PR) of
the sample.  This process is impeded by the pinning forces encountered
when a vortex enters the PR. As more vortices are added,  other
vortices are gradually pushed farther into the PR, in both directions,
until the two developing ``flux fronts" meet in the middle of the PR.
This situation corresponds to reaching the full penetration field $H^*$
discussed in Bean's work \cite{bean}.  We then continue  to add vortices 
(up to 1,600) until the system is quite densely packed.  The results of one
such simulation are shown in Fig.~1a.  In this figure, the circular
sample is cut in the middle of the UR and unfolded for visualization
purposes.  The plot is of vortex density (which is
proportional to field) versus distance across the sample.  One can
clearly see the development of Bean-like flux profiles, where the flux
density is highest on the edges of the sample and lowest in the
middle.  The density in the UR can be thought of as the ``external"
field and the design of the simulation is such that the boundary
condition of the internal field at the edges of the sample being equal
to the external field is automatically met.

Starting from the configuration that results upon the completion of the
ramp-up phase,  we then successively remove vortices from the UR and
allow the system to relax as before.  This procedure is continued until
all vortices are removed from the UR.  The configuration at this point
corresponds to the remnant magnetization peak which results when an
external field is raised to some value and then brought back down to
zero.  The results of a ramp-down simulation are shown in fig. 1b.
Again, Bean-like behavior is evident, including the characteristic
``gull-wing" shaped profiles which result when the memory of the
initial configuration before the ramp-down is not yet erased.

Exploiting the analogy between density in the UR and external field,
the results can be used to calculate a magnetization hysteresis loop
for the sample. In actuality, we create a partial loop since our
simulation does not readily allow for a reversal of field direction.
The density in the UR is taken to be the external field and the
integrated difference between this field and the internal field across
the sample yields the magnetization.  This result is illustrated in
Fig.~2 for two different pin densities.

An interesting feature of the profiles shown in Fig.~1 is that one can
observe a clear trend in the slope of the flux density as a function of
field.  This slope corresponds to the critical current $J_{\rm c}$ in
real samples, and it can be seen to decrease as the local field
increases.  This variation in $J_{\rm c}$ leads to curved profiles,
rather than the linear profiles in the classical Bean picture.  Such a
variation is often described as ``modified" Bean behavior and several
forms of modification have been previously proposed.

One of the earliest of these was put forward in Ref.\cite{kim} shortly
after Bean's original work.  There, the authors define a maximum
pinning force $\alpha_m$ and find empirically that this maximum force
does not vary with field.  The critical current at a given field is
then found by equating the Lorentz force with $\alpha_m$ as in the
following:
\begin{equation}
\alpha_m = J_{\rm c}(H + B_0),
\end{equation}
Where $\alpha_m$ and $B_0$ are constants which depend on the
microstructure of the material and $H$ is the local field.  This form
implies that the inverse of $J_{\rm c}$ should depend linearly on
field.

By performing linear fits to  portions of profiles like those shown in
Fig.~1 over a range of fields,  we are able to extract the dependence
of $J_{\rm c}$ on field in our samples. The inverse of $J_{\rm c}$ for
one simulation is shown in Fig.~3. One can see that a linear dependence
of this quantity on field describes the data very well.  One can also
observe from the plot an apparent oscillatory behavior superimposed on
the linear trend in $1/J_{\rm c}$.  This is an artifact of the finite
range of the vortex force, which results in a small discontinuity in
the slope whenever the number of vortices within the force range of a
given vortex increases by one.  In a real sample, this change would
happen gradually.

To further explore the relationship of our data to the Kim model of
Ref. \cite{kim},
 we examined the dependence of the maximum pinning force $\alpha_m$ on
 pin density  and the vortex-vortex force.  Varying the latter quantity
is equivalent to varying the pinning strength since it is the ratio of
the pinning strength (defined as one) to ${\rm A}_{\rm v}$ which
determines the behavior.  The results are shown in the insets in
Fig.~3.  As expected, $\alpha_m$ increases monotonically with the
number of pinning centers and with the pinning strength, $1/{\rm
A}_{\rm v}$.

We also investigated the dynamics of the system after a single vortex
has been added or subtracted.  This was accomplished by evaluating the
displacements of all the vortices in the system after each step.  From
this information, one can extract most dynamical quantities of
interest.  In particular, one can get an idea of the magnitude of the
global motion of the system after a vortex addition or subtraction by
calculating the average displacement of all of the vortices in the
system at that step.
 Examination of this quantity is complicated by the fact that the
 properties of the system evolve as the vortex density changes.  The
 individual displacements of vortices, and hence the global average,
 are generally smaller when the system is more highly filled since the
 motion of each vortex is constricted by the close proximity of its
 neighbors.  The simplest solution of examining only a small portion of
 the simulation over which the properties of the system do not vary
 much, places too severe a restriction on the statistics for this
 quantity.  Our approach, then, consisted of normalizing the average
 displacement at each step according to that which would be expected
 for an unpinned periodic system with the same number of vortices.  The
 average displacement for a vortex in an unpinned system can be shown
 to depend on total vortex number $N$ as $L/4N$, where $L$ is the
 length of the system.

By dividing the average displacement at each step by this normalizing
function, we were able to eliminate any systematic trends in the data
over the course of the simulation and thereby focus upon the
statistical fluctations from step to step. The distibution of these
normalized average displacements is shown in Fig.~4 for a simulation
with a pin density of 20 per unit length, and $A_{\rm v} = 5$. The
ramp-up and ramp-down data were combined in this plot since, when
examined alone, their distributions were comparable. We see power-law
behavior over approximately 2.5 decades, with a slope of 0.71.

In addition, we examined the distribution of displacements of
individual vortices in the system.  Though this quantity is more
difficult to normalize, we see  power law behavior over approximately 
2.5 decades here as well.

The theoretical treatment of the Bean model usually involves discussion
of the Lorentz force on vortices.  This force is proportional to the
product of the local field and the current at that point.  As
emphasized by Brandt \cite{brandt1}, the only way a macroscopic
transport current can exist in a type-II superconductor is if it is
accompanied by a gradient in the density of vortices or curvature of
vortices.  This is often a point of confusion in work on the subject of
magnetic and transport properties of superconductors.  In our
simulation, no Lorentz force, per se, is applied to the vortices in the
sample.  They are simply driven into the sample by the repulsive forces
between vortices. This repulsion, in conjunction with the interaction
with attractive pinning centers, leads naturally to the development of
a gradient in the vortex
density, and vortices tend to move down this gradient.  Now in a real
sample, this gradient would necessarily, by Maxwell's equations, be
accompanied by a current and one could describe the force on each
vortex as being a Lorentz force which results from the interaction of
current and magnetic field.  Our work emphasizes the fundamental
equivalence of these two outlooks, {\em i.e.,} one can equally well
describe the force on a vortex as being due to a Lorentz type process
or as resulting from a greater repulsive force from the side of the
vortex where there is a higher density of neighbors.

Another important result from our simulations is the natural
development of a systematic decrease in $J_{\rm c}$ with increasing
field.  We stress again that the physical input into these simulations
is very simple, consisting only of linear attractive and repulsive
forces.   That this is sufficient to derive a dependence of $J_{\rm c}$
on field that is consistent with experiment suggests that one need look
no further to explain this behavior than that the increase in repulsive
forces that accompanies an increase in field makes the pinning force,
which does not vary with field, gradually less important.

>From a dynamic point of view, there has been considerable interest and
speculation about systems of the type under investigation here possibly
exhibiting self-organized criticality (SOC) when driven to the
threshold of instability.  Many discussions of SOC in pinned flux
lattices have relied on measurements or investigations of systems {\em
relaxing away} from criticality \cite{vinokur,ling}, making them less
than ideally suited for examining the SOC hypothesis.  In our work, the
system is truly driven to the threshold of instability and then allowed
to organize itself into a critical state.  In this manner, our
simulations are analogous to slowly dropping sand on the top of a pile
and observing the subsequent avalanches.  It is generally believed that
the strongest indicator of a system having an SOC character is for the
distribution of dynamical events associated with that system to obey a
power law.
The average vortex displacements, shown in Fig.~4, are a measure of the
overall ``avalanche" activity in our systems at a given step.  That
this distribution follows a power law  over more than two decades
suggests that these systems may indeed be SOC in nature.

The authors would like to acknowledge useful discussions with S. Field
and J. Witt. FN acknowledges partial support from a GE fellowship, a
Rackham grant, the NSF through grant DMR-90-01502, and SUN
Microsystems. OP is supported by a fellowship from Ministerio de
Educaci\'on y Ciencia, Spain and by DoE grant DE-FG-02-85ER5418.

\begin{figure}
\caption{Vortex density profiles for the (a) ramp-up phase and (b)
ramp-down phase for a simulation with a pin density of 30 per unit
length and ${\rm A}_{\rm v}=5$. The flat plateaus on either side of the
sample show the density in the unpinned region  and the jagged V-shaped
profiles correspond to the density in the pinned region.}
\end{figure}

\begin{figure}
\caption{Magnetization curves for two simulations.  The outer, darker
curve is for a pin density of 30 and the inner curve is for a pin
density of 15.  Both samples have ${\rm A}_{\rm v}=5$.  The x-axis is
the vortex density in the unpinned region (UR), which corresponds to
external field.}
\end{figure}

\begin{figure}
\caption{1/$J_{\rm c}$ for a sample with a pin density of 20 and ${\rm
A}_{\rm v}=5$.  Solid line is a linear fit to the data.  Insets show
the dependence of the maximum pinning force $\alpha_m$ on $1/{\rm
A}_{\rm v}$, corresponding to pinning strength, and pin density. The
values of $\alpha_m$ are obtained from linear fits like that shown in
the main figure.}
\end{figure}

\begin{figure}
\caption{Distribution of normalized average displacements for a
simulation with pin density 20 and ${\rm A}_{\rm v}=5$. The normalized
average displacement, d, is plotted on the horizontal axis and D(d),
proportional to the number of events with a given d, is plotted on the
vertical axis.}
\end{figure}


\begin{references}
\bibitem[\dag]{dagger} Author to whom correspondence should be addressed.
\end{references}
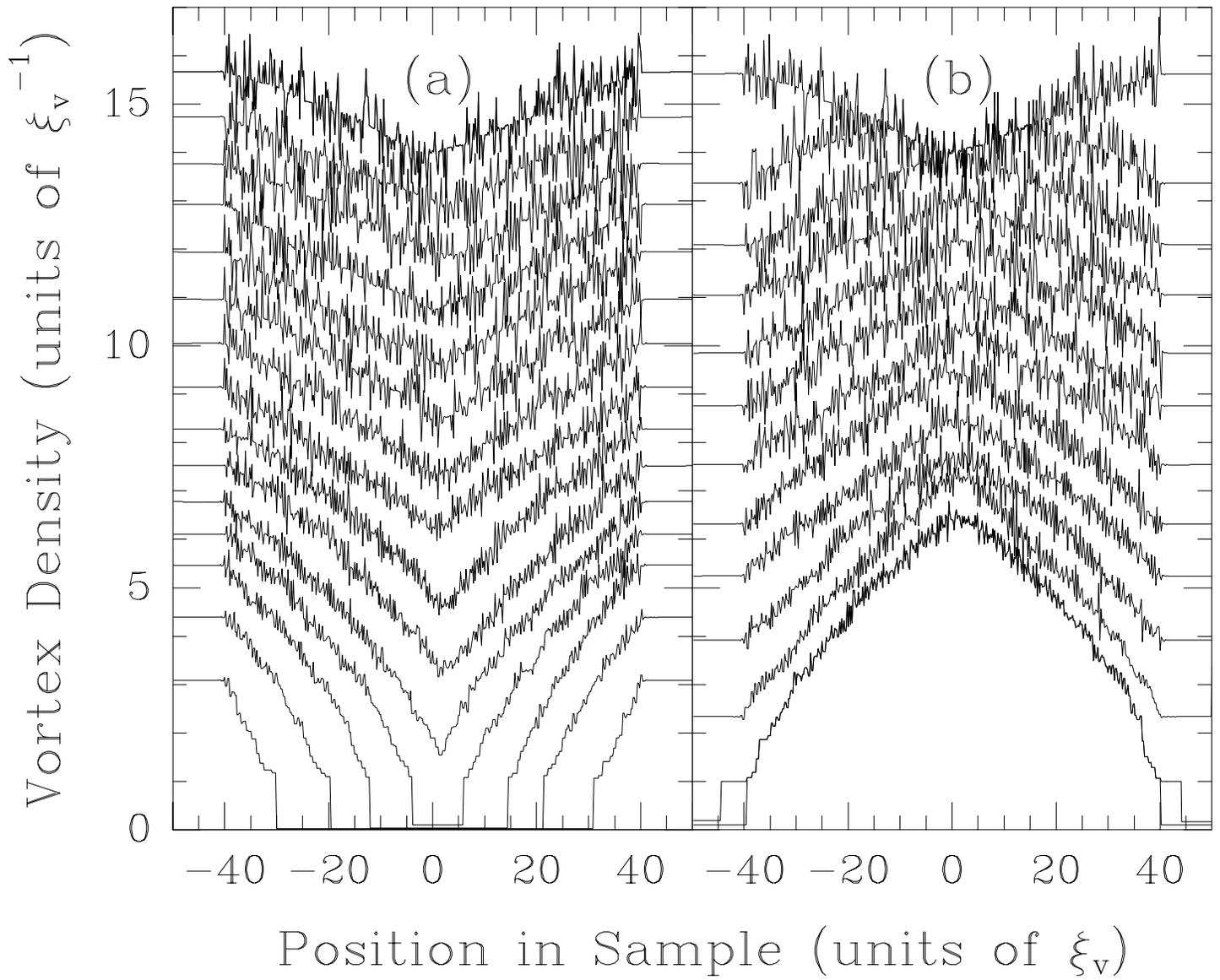
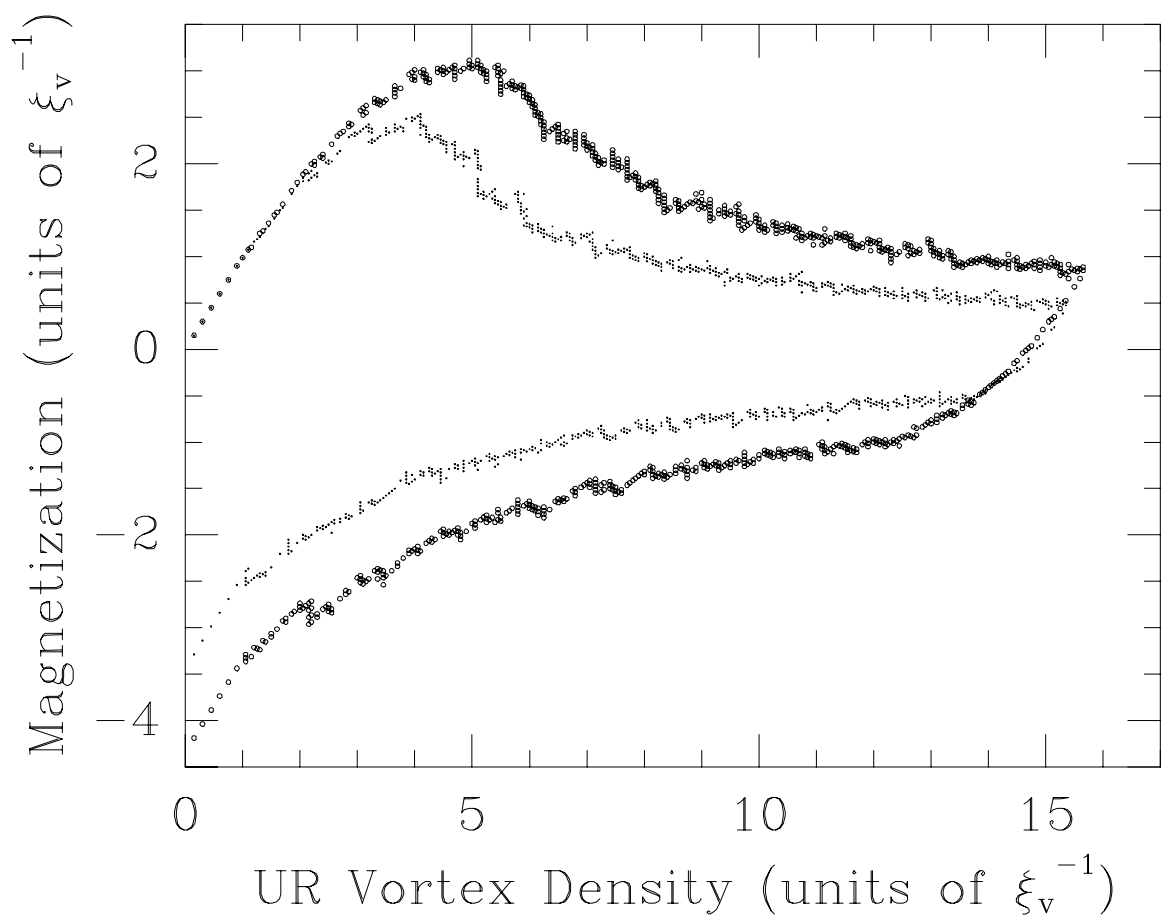
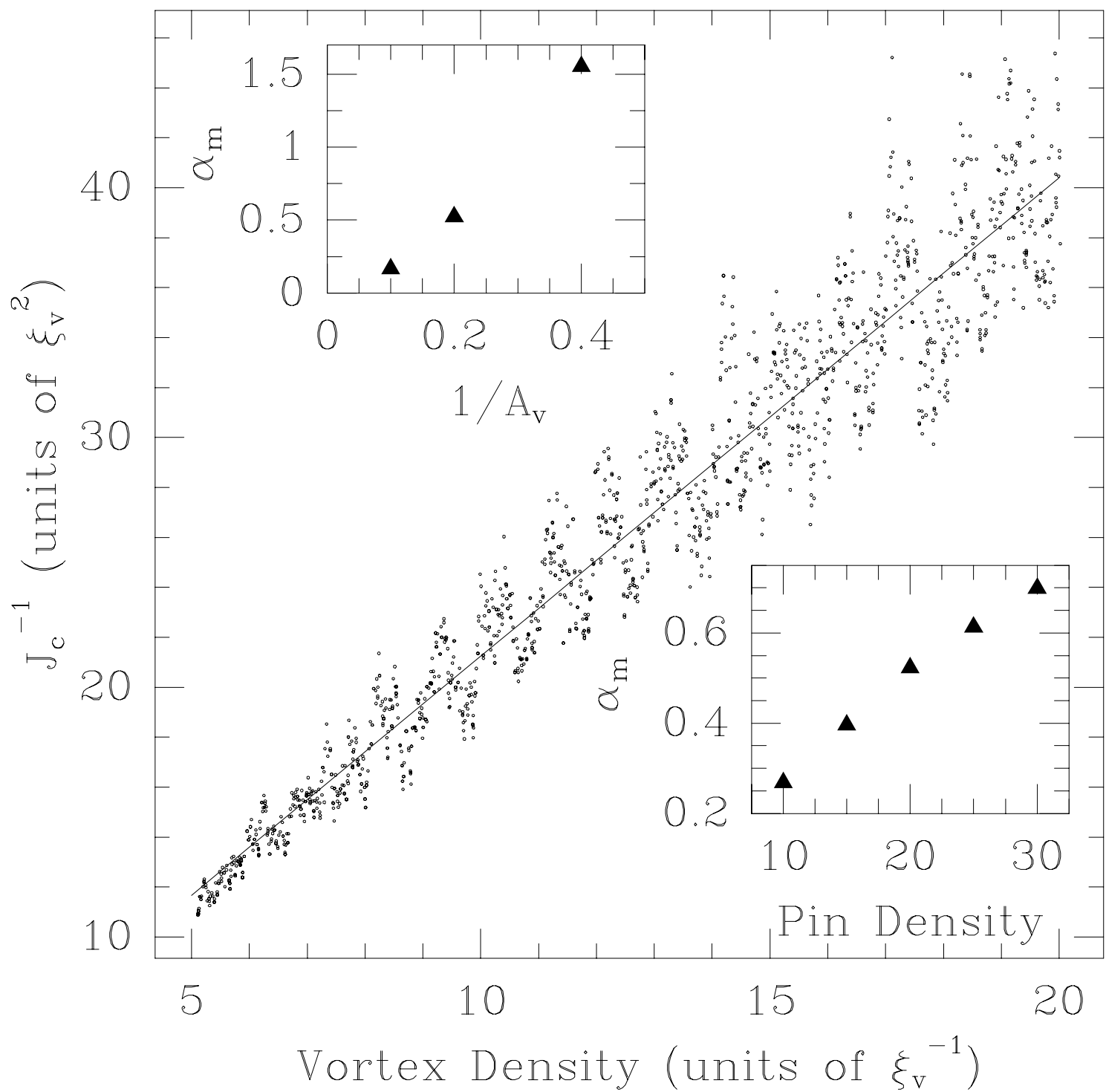
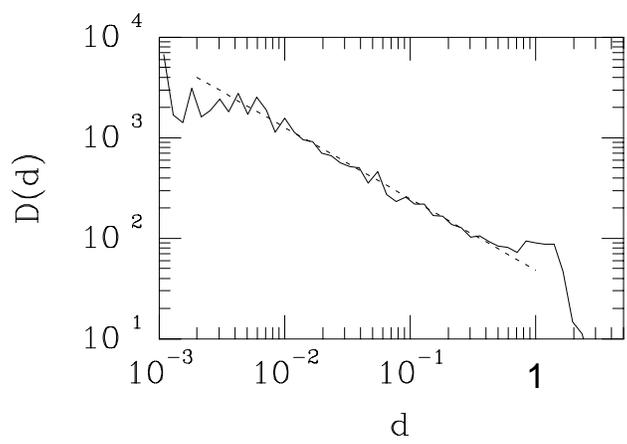
\end{document}